\documentclass[apj, paper]{emulateapj}
\pdfoutput=1
\usepackage{threeparttable}
\usepackage{amsmath,bm}
\usepackage{gensymb}
\usepackage{multirow}
\usepackage{lineno}
\usepackage{color} 
\usepackage{xcolor}
\shorttitle{A new LBV in M31}
\shortauthors{Y. Huang et al.}

\begin{document}
\slugcomment{\bf}

\title{A New Luminous blue variable in the outskirt of the Andromeda Galaxy}

\author{Y. Huang\altaffilmark{1}}
\author{ H.-W. Zhang\altaffilmark{2,3}}
\author{ C. Wang\altaffilmark{2,10}}
\author{ B.-Q. Chen\altaffilmark{1}}
\author{Y.-W. Zhang\altaffilmark{4,5}}
\author{J.-C. Guo\altaffilmark{2,9}}
\author{ H.-B. Yuan\altaffilmark{6}}
\author{ M.-S. Xiang\altaffilmark{7}}
\author{ Z.-J. Tian\altaffilmark{8}}
\author{ G.-X. Li\altaffilmark{1}}
\author{ X.-W. Liu\altaffilmark{1}}

\altaffiltext{1}{South-Western Institute for Astronomy Research (SWIFAR), Yunnan University (YNU), Kunming 650500, People's Republic of China; {\it yanghuang@ynu.edu.cn;  x.liu@ynu.edu.cn}}
\altaffiltext{2}{Department of Astronomy, Peking University (PKU), Beijing, 100871, People's Republic of China; {\it zhanghw@pku.edu.cn}}
\altaffiltext{3}{Kavli Institute for Astronomy and Astrophysics, Peking University, Beijing, 100871, People's Republic of China}
\altaffiltext{4}{University of Chinese Academy of Sciences, Beijing 100049, People's Republic of China}
\altaffiltext{5}{Yunnan Observatories, Chinese Academy of Sciences, Kunming 650011, People's Republic of China}
\altaffiltext{6}{Department of Astronomy, Beijing Normal University, Beijing, 100875, People's Republic of China}
\altaffiltext{7}{Max-Planck Institute for Astronomy, K{\"o}nigstuhl, D-69117, Heidelberg, Germany}
\altaffiltext{8}{Department of Astronomy, Yunnan University (YNU), Kunming 650500, People's Republic of China}
\altaffiltext{9}{Department of Physics and Astronomy, University College London, London WC1E 6BT, UK}
\altaffiltext{10}{LAMOST Fellow}
\altaffiltext{11}{YNU and PKU are the co-first unit.}

\begin{abstract}
The hot massive luminous blue variables (LBVs) represent an important evolutionary phase of massive stars.
Here, we report the discovery of a new LBV -- LAMOST\,J0037+4016 in the distant outskirt of the Andromeda galaxy. 
It is located in the south-western corner (a possible faint spiral arm) of M31 with an unexpectedly large projection distance of $\sim$ 22\,kpc from the center.
The optical light curve {shows a 1.2 mag variation} in $V$ band and its outburst and quiescence phases both last over several years.
The observed spectra indicate an A-type supergiant at epoch close to the outburst phase and a hot B-type supergiant with weak [Fe II] emission lines at epoch of much dimmer brightness.
The near-infrared color-color diagram further shows it follows the distribution of Galactic and M31 LBVs {rather than B[e] supergiants}.
All the existing data strongly show that  LAMOST\,J0037+4016 is an LBV.
By spectral energy distribution fitting, we find it has a luminosity ($4.42 \pm 1.64$)$\times 10^5$\,$L_{\odot}$ and an initial mass $\sim 30$\,$M_{\odot}$, indicating its nature of less luminosity class of LBV.
\end{abstract}
\keywords{galaxies: individual (M31) -- stars: massive -- supergiants}

\section{Introduction}
Luminous blue variables (LBVs) are hot, unstable, massive ($\ge 25$-$30$\,$M_{\odot}$) and extremely luminous ($10^{5}$--$10^{7} L_{\odot}$) evolved stars in the upper left parts of the Hertzsprung-Russell (HR) diagram (Conti 1984; Humphreys \& Davidson 1994, hereafter HD94).
During the LBV phase, massive stars undergo episodes of eruptive mass-loss ($10^{-5}$--$10^{-4} M_{\odot} {\rm yr}^{-1}$), accompanied by spectacular photometric and spectral variabilities, on timescales of years to decades or longer (HD94; van Genderen 2001).
The origin of this eruption is not yet well understood, partly due to the limited number of confirmed LBVs and the infrequency of the eruption stage. 
From the quiescence to the outburst (or eruptive) stage, the brightness of  an LBV increases by $1$--$2$\,mag in the visual band and the spectrum changes from an O/early B-type supergiant to an A/F-type one.

LBVs are of vital importance for studying the very{ late stage of massive stars evolution}.
However, the current understanding of the evolutionary status of LBVs is quite uncertain.
At present, the LBV is either considered as a transition phase between the main sequence of massive
stars and Wolf-Rayet (WR) stars (e.g. HD94; Maeder \& Meynet 2000) or is an intermediate precursor of a supernova (e.g. Smith et al. 2007, 2008, 2011; Trundle et al. 2008; Gal-Yam \& Leonard 2009; Groh, Meynet \& Ekstr{\"o}m 2013; Mauerhan et al. 2013; see more details in the review paper by Smith 2014).
More recently, Smith \& Tombleson (2015) propose that LBVs might be evolved blue stragglers, products of binary evolution, although {this is} still on hot debate (e.g. Humphreys et al. 2016; Davidson et al 2016; Aadland et al. 2018).
Discovery of new LBVs is therefore of vital importance for understanding the origin of the eruption and the role LBVs play in the stellar evolution.

In this letter, we report the discovery of {a new LBV} -- LAMOST\,J0037+4016 (right ascension: 00:37:20.65, declination: $+$40:16:37.70) in the Andromeda galaxy (M31).
The object was originally selected as an M31 blue supergiant candidate for LAMOST (Cui et al. 2012) observation   based on its Johnson $Q$-index (e.g. Massey et al. 1998) calculated using the photometry from the Local Group Galaxy Survey (LGGS; Massey et al. 2006).
The resultant radial velocity agrees very well with what expected as predicted by the rotation curve (Drout et al. 2009) of M31, and thus confirms it is indeed one of supergiants in M31.
This supergiant is in the south-western corner of M31 at an unexpectedly large projection distance of $\sim 22$\,kpc from the centre of M31 (see Fig.\,1).
This corner seems belong to a extended (sub-)structure (possibly a faint spiral arm) of the M31 disc, as revealed by the {\it Herschel} SPIRE $250$\,$\mu$m (Fritz et al. 2012) and 21\,cm (Braun et al. 2009) images shown in Fig.\,1.
In Section\,2, we describe the photometric and spectroscopic
data that we employ in the current work. 
The main results are presented and discussed in Section\,3. 
In Section\,4, the physical properties (e.g. effective temperature, luminosity) and environments of LAMOST\,J0037+4016 are derived and discussed{, respectively}. 
Finally, we summarize in Section 5.

\begin{figure*}
\begin{center}
\includegraphics[scale=0.45,angle=0]{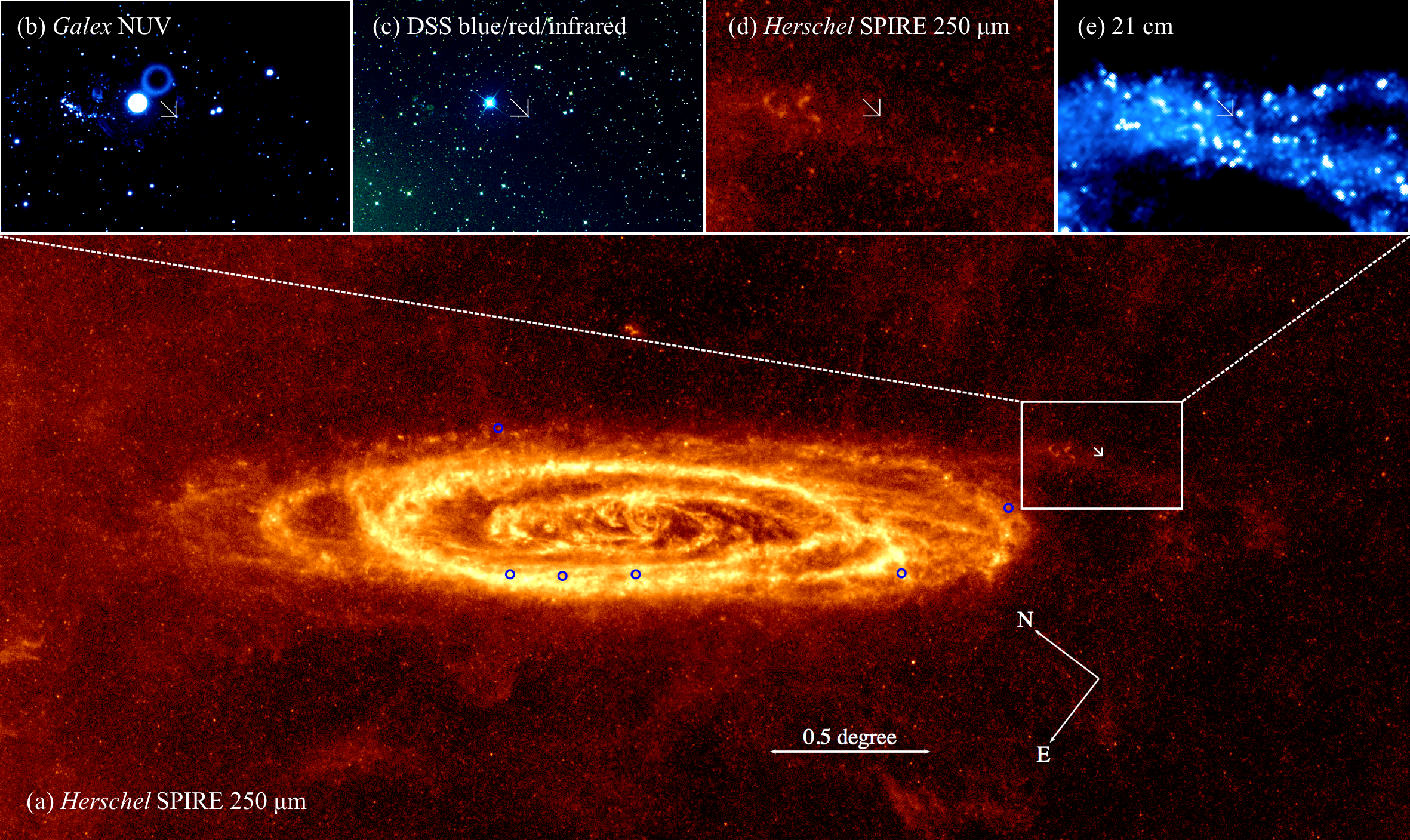}
\caption{Images in the different bands to show the location and environment of LAMOST\,J0037+4016. Panel\,a: {\it Herschel} SPIRE 250\,$\mu$m image.
              Spatial scale is shown in the middle-bottom. The six blue circles mark the locations of the confirmed LBVs (Richardson \& Mehner 2018) in M31.
              The white box with sizes of $20$\arcmin\,$\times 30$\arcmin\,is centred at the location of this newly discovered LBV.
              The four images on the top zoom the area (of the same size of the white box) of LAMOST\,J0037+4016 in {\it Galex} NUV band (panel\,b); in DSS blue, red and infrared color composite image (panel\,c); in  {\it Herschel} SPIRE 250\,$\mu$m image (panel\,d) and in 21\,cm image (panel\,e).}
\end{center}
\end{figure*}
\newpage
\section{Data}
To unravel the nature of LAMOST\,J0037+4016, we first describe the available photometric and spectroscopic data, including both archival and newly obtained in the current work.

\subsection{Photometric data}
The photometric data of LAMOST\,J0037+4016 employed here are mainly from the Catalina Real Time Transient Survey (CRTS; Drake et al. 2009) and the All-Sky Automated Survey for Supernovae (ASAS-SN; Jayasinghe et al. 2019).
From CRTS, measurements of a total of 207 epochs in $V$ band are found for LAMOST\,J0037+4016 over a period of $\sim 8$ years from 2005 to 2013. 
All these measurements were collected with the 0.7\,m Catalina Schmidt Survey (CSS) telescope located north of Tucson, Arizona.
From ASAS-SN, a total of 741 measurements in Johnson $V$ band at different epochs are found for LAMOST\,J0037+4016 over a period of $\sim 7$ years from 2012 to 2018.
We note only 294 of the 741 measurements with magnitude errors smaller than 0.15\,mag are finally adopted in the current analysis.

In addition to the CRTS and ASAS-SN data, we have also included some observations from large-scale multi-band optical/near-infrared photometric surveys, including LGGS ($UBVRI$; Massey et al. 2006), SDSS ($ugriz$; Alam et al. 2015); XSTP-GAC ($gri$; Liu et al. 2014; Zhang et al. 2014) and 2MASS ($JHK_{\rm s}$; Skrutskie et al. 2006).

Finally, we observed LAMOST\,J0037+4016 with the 2.16\,m telescope of NAOC in Xinglong (XL), China,  in $B$ and $V$ bands on 2016 November 7 and with the 2.4\,m telescope of YNAO in {Lijiang (LJ)}, China, in $UBVRI$ bands on 2017 January 13.
The data were reduced in the standard way with IDL programming language. 
The resultant magnitudes were then calibrated to the Johnson-Cousins photometric systems.
All the photometric observations described above are summarized in Table\,1.

\subsection{Spectroscopic data}
As mentioned, LAMOST\,J0037+4016 was selected and targeted by the LAMOST Spectroscopic Survey (Deng et al. 2012; Liu et al. 2014) as a supergiant candidate in M31.
It was observed twice by LAMOST, on 2011 October 4 and on 2011 October 29. 
LAMOST can simultaneously collect up to 4000 optical ($\lambda\lambda$ 3700-9000), low resolution ($R \sim 1800$) spectra in one exposure. 
The data were processed using the LAMOST standard pipeline (Xiang et al. 2015).
In addition, LAMOST\,J0037+4016 was observed on 2016 January 26 and  on 2018 January 22 with YFOSC mounted on the 2.4\,m telescope of YNAO  and with the double spectrograph (DBSP) mounted on 5.1\,m Hale telescope.
YFOSC is a multi-modes instrument for imaging and low/medium resolution spectroscopy, working at the Cassegrain focus.
A slit width of 1\farcs8 was used with grism\,14 (G14), covering 3700-7200\,\AA\,at a full width at half maximum (FWHM) resolution of $\sim 11.4$\,\AA.
The DBSP is a low-to-medium resolution ($R$ between 1000 and 10000) grating instrument and it can obtain spectra in blue and red channels ({using a beam splitter}), simultaneously.
Using a slit of 1\arcsec, together with B1200 and R600, we have obtained a blue spectrum covering 3890-5400\,\AA\,({FWHM  $\sim$\,1.4\,\AA}) and a red spectrum covering 5790-9000\,\AA\,({FWHM  $\sim$\,2.8\,\AA}).
The data taken from YFOSC and DBSP were reduced with the standard procedure using IRAF.
The details of those spectroscopic observations are summarized in Table\,2. 

\begin{table}
\caption{Photometric data used for LAMOST\,J0037+4016}
\centering
\begin{tabular}{lccc}
\hline
UT date&Facility&Filters&Number of epochs\\
\hline
1998 Nov. 23&2MASS&$JHK_{\rm s}$&1\\

2000 Oct. 04&LGGS&$UBVR$&1\\

2002 Sep. 05&SDSS&$ugriz$&1\\

2005 -- 2013&CRTS-CSS&$V$&207\\

2010 Oct. 01&XSTP-GAC&$gri$&1\\

2012 -- 2018&ASAS-SN&$V$&294\\

2016 Nov. 07&NAOC 2.16\,m&$BV$&1\\

2017 Jan. 13&YNAO 2.4\,m&$UBVRI$&1\\
\hline
\end{tabular}
\end{table}

\begin{table*}
\caption{Spectroscopic data used for LAMOST\,J0037+4016}
\centering
\begin{tabular}{lccccc}
\hline
UT date&Facility&Wavelength range&FWHM& Exp. time&Seeing\\
&&(\AA)&(\AA)&(s)&(arcsec)\\
\hline
2011 Oct. 04&LAMOST&3700-9000&2.5&1800$\times$3&3.2\\

2011 Oct. 29&LAMOST&3700-9000&2.5&900$\times$3&2.8\\

2016 Jan. 26&YNAO 2.4\,m/YFOSC-G14&3700-7200&11.4&1800&1.3\\

\multirow{2}{*}{2018 Jan. 22}&Palomar 5.1\,m/DBSP-B1200&3890-5400&1.4&\multirow{2}{*}{1200$\times$3}&\multirow{2}{*}{1.1}\\

&Palomar 5.1\,m/DBSP-R600&5790-9000&2.8&&\\

\hline
\end{tabular}
\end{table*}

\section{Results}
\subsection{Light curve}
To construct the optical light curve of LAMOST\,J0037+4016, we first convert measurements in other systems  to Johnson $V$ band magnitudes.
For the SDSS filter system, we obtain Johnson $V$ magnitudes using the transformation equation of Jester et al. (2005),
\begin{equation}
V = g - 0.58 \times (g - r) - 0.01\text{.}
\end{equation}
For the CRTS-CSS, Johnson $V$ band magnitudes are calculated using the transformation equation of Graham et al. (2015),
\begin{equation}
V = V_{\rm CSS} + 0.31 \times (B-V)^2 + 0.04\text{,}
\end{equation}
where color $B-V$ is adopted as the mean value of measurements given by LGGS, SDSS and XSTP-GAC\footnote{For the latter two surveys, values of color $B-V$ are actually those converted from color $g-r$ (Jester et al. 2005)}.
The measurements obtained by LGGS, ASAS-SN, NAOC 2.16\,m and YNAO 2.4\,m are already of the Johnson system.

The resultant $V$ band light curve of LAMOST\,J0037+4016 over a period of $\ge 18$ years is presented in Fig.\,2(a).
The plot shows that LAMOST\,J0037+4016 continued to rise in brightness since the beginning of the data, 2000 October ($V = 15.85$\,mag), and reached maximum light (the outburst phase) around 2005 September ($V \sim$\,15.2-15.4\,mag).
The outburst phase lasted about five years. 
It started to decline from 2011 January and reached minimum light (the quiescence phase) around 2015 January ($V \sim$\,16.25-16.50\,mag).
Until the last observation at the end of 2018 November,  LAMOST\,J0037+4016 was still at the quiescence phase.
Here we note that the large variations within few to tens of days from ASAS-SN photometry are largely contributed from their measurement errors.
The current light curve shows a $V$ band amplitude ($\Delta V$) over $1.2$\,mag and an outburst/quiescence phase lasting several years.
Both behaviours suggests that LAMOST\,J0037+4016 is a typical LBV (HD94).

\begin{figure*}
\begin{center}
\includegraphics[scale=0.52,angle=0]{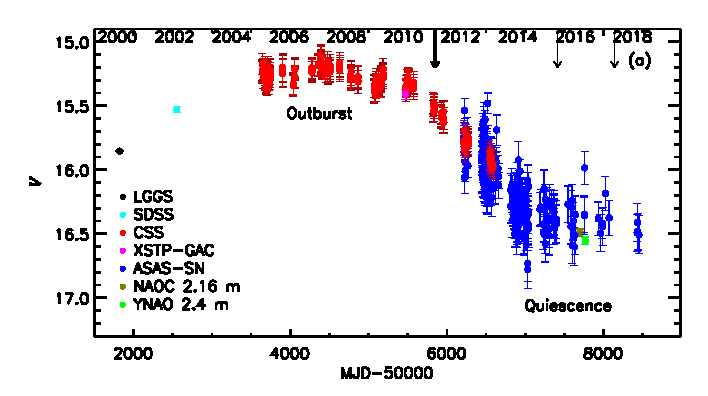}
\includegraphics[scale=0.52,angle=0]{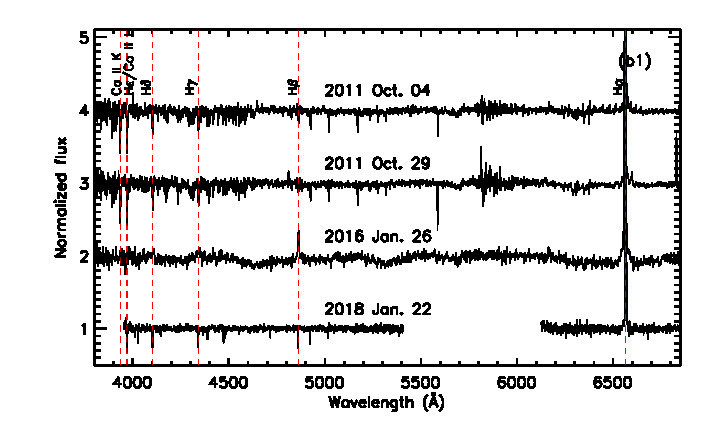}
\includegraphics[scale=0.30,angle=0]{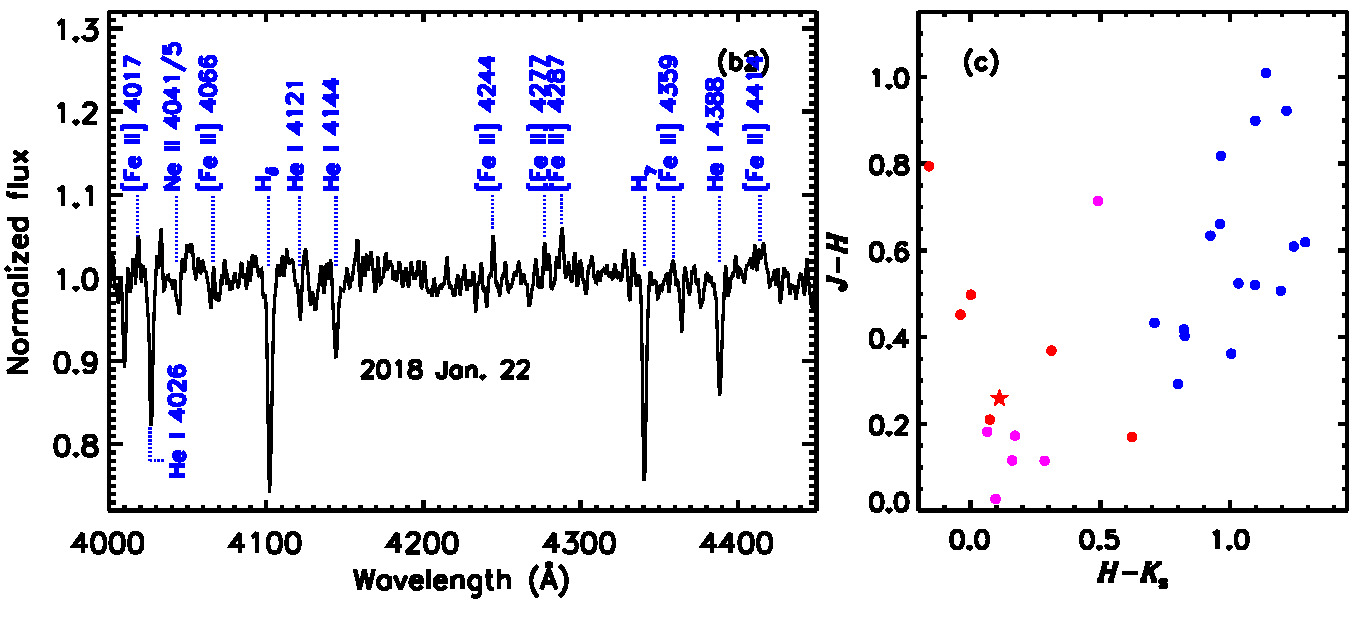}
\caption{Optical light curve, spectra and near infrared colour-colour diagram of LAMOST\,J0037+4016. {a,} $V$ band light curve of LAMOST\,J0037+4016 (see Section\,3.1).
              The four arrows mark the dates of the four spectroscopic observations of this target. {b1,} Normalized spectra of LAMOST\,J0037+4016. The Balmer and Ca\,{\sc ii} lines are identified.
              {b2,} The blue range of the spectrum taken at 2018 January 22. The Balmer, He\,{\sc i} and some weak [Fe {\sc ii}] lines are identified.
              {c,} $J - H$ versus $H - K_{\rm s}$ colour-colour diagram of LBVs and B[e]SGs. Galactic LBVs and B[e]SGs are shown by magenta and blue filled circles, respectively. The six confirmed M31 LBVs are shown by red filled circles.
              The red filled star represents LAMOST\,J0037+4016. }
\end{center}
\end{figure*}

\begin{figure*}
\begin{center}
\includegraphics[scale=0.44,angle=0]{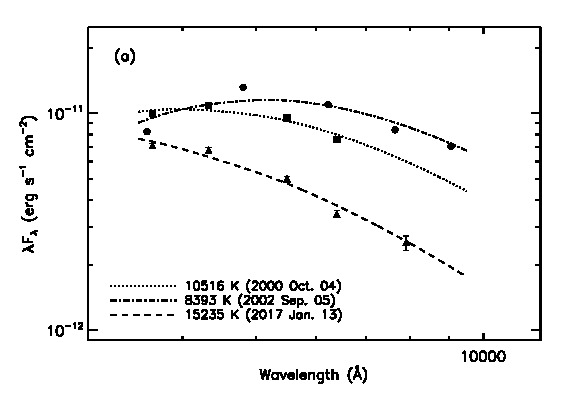}
\includegraphics[scale=0.44,angle=0]{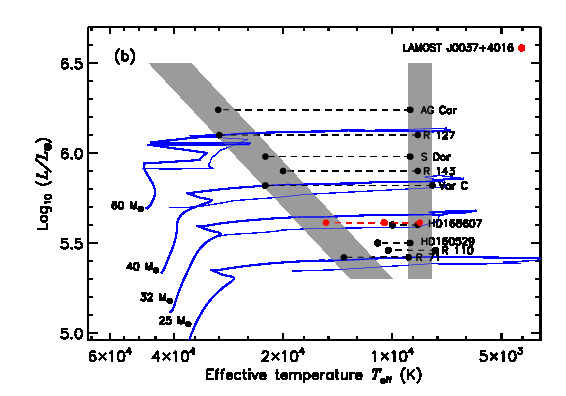}
\caption{ {\it Left panel:} SED fitting for LAMOST\,0037+4016.
               Squares, dots and triangles represent the multi-bands ($UBVRI$/$ugriz$) measurements of LGGS, SDSS and YNAO 2.4 m, respectively.
               The dash-dot, dot and dash lines show the best fits of SEDs based on data from LGGS, SDSS and YNAO\,2.4\,m, respectively.
                The best-fit effective temperatures are labelled at the bottom-left corner of the plot.
                {\it Right panel:} HR diagram of LBVs. 
                The right shaded area represents the outburst phase with a nearly constant effective temperature.
                The left shaded area represents the diagonal S\,Doradus instability strip adopted from HD94.
                Black dots represent well-known LBVs with values of $L$ and $T_{\rm eff}$ adopted from HD94 and de Jager (1998).
                Red dots represent LAMOST\,0037+4016 with values of $L$ and $T_{\rm eff}$ derived in {Section 4.1}.
                Also overplotted blue lines are Geneva evolutionary tracks (Ekstr{\"o}m et al. 2012) of solar metallicty and an initial rotation speed of 40\% of the breakup speed.
                Here we only include the first 209 points, representing the beginning of any blue loop, for simplicity.
               }
\end{center}
\end{figure*}

\subsection{Spectra}
In addition to the light curve, we have obtained four spectra at different phases of LAMOST\,J0037+4016.
As shown in Fig.\,2(b1), the top two spectra were taken by LAMOST on the time (4 and 29 October, 2011) close to the outburst phase of LAMOST\,J0037+4016.
The two spectra are almost identical, given the small time interval of less than a month.
A consistent radial velocity of $-488.7$\,km\,s$^{-1}$ is derived from both spectra and it agrees very well with what expected ($-497.7$\,km\,s$^{-1}$), as predicted by the rotation curve of M31 (Drout et al. 2009), confirming that LAMOST\,J0037+4016 belongs to M31.
We also note the presence of H$\alpha$ emissions in both spectra.
The third spectrum of LAMOST\,J0037+4016 was obtained on 2016 January  (the quiescence phase) using YFOSC+G14 mounted on Yunnan Observatories (YNAO) 2.4\,m with a 1\farcs8 wide slit.
The spectral resolution is quite low, $R \sim 500$.
The spectrum is plotted in Fig.\,2(b1) and shows signicant variations in Balmer lines (i.e. H$\alpha$, H$\beta$, H$\gamma$ and H$\delta$) compared to the earlier two LAMOST spectra.
H$\alpha$ became much stronger this time and the other three high-order Balmer lines all turned from absorption to emission.
The last spectrum of LAMOST\,J0037+4016 was obtained on 2018 January (the quiescence phase) using the double spectrograph (DBSP) mounted on 5.1\,m Hale telescope.
The spectrum is shown in Fig.\,2(b1).
The emissions of H$\beta$, H$\gamma$ and H$\delta$ turned to absorptions again.
Specially, we zoom the blue range of this spectrum in Fig.\,2(b2) and some weak [Fe II] emission lines are clearly seen.

We further  classify the spectral and luminosity type of LAMOST\,J0037+4016 by comapring the obtained spectra to a collected spectral library using template matching with a least $\chi$-square algorithm.
The collected library includes 160 spectra (FWHM$\sim 1.8$\AA) of O type stars from the Galactic O-star catalog (Sota et al. 2011), 40 spectra (FWHM$\sim 2.0$\AA) of O/B type stars from the digital atlas of OB stars (Walborn et al. 1990), and 70 spectra (FWHM$\sim 2.5$\AA) of B/A type stars from the MILES spectral library (S{\'a}nchez-Bl{\'a}zquez et al. 2006).
To provide a uniform spectral library, a smoothing filter was applied to all the collected spectra to achieve a constant FWHM\,$= 2.5$\,\AA\,for a common spectral range 4000-4820\,\AA.
When compared to the spectral library, the observed spectra are also degraded to FWHM\,$= 2.5$\,\AA.
For the two spectra obtained by LAMOST, they are all classified as A3Ia of an effective temperature around 9000\,K, consistent with the expected type/temperature at their observational epochs.
We do not obtain the spectral and luminosity type for the third  due to its low spectral resolution.
For the last spectrum, It is classified as B5/8Ia of an effective temperature around 12000\,K from the template matching.
From those {spectra}, we find that LAMOST\,J0037+4016 {transited} from a cool to a warm state from 2011 October to 2018 January, in excellent agreement with the variations seen in the optical light curve.

\begin{table*}
\caption{Stars within 20\arcsec of LAMOST\,J0037+4016 with $UBV$ photometry from the LGGS}
\centering
\begin{tabular}{ccccc}
\hline
Star Name& $V$ & $B-V$ & $U-B$ & $Q$\\
&(mag) & (mag) & (mag) &\\
\hline 
J003719.71+401634.3 & $      22.32\pm     0.12$ & $    -0.13\pm     0.15$ & $    -0.79\pm    0.10$ & $    -0.70\pm     0.14$\\
J003719.97+401627.7 & $      22.15\pm    0.08$ & $    -0.23\pm     0.10$ & $    -0.97\pm    0.07$ & $    -0.80\pm    0.10$\\
J003720.04+401635.3 & $      22.00\pm    0.09$ & $    0.03\pm    0.10$ & $     -1.05\pm    0.05$ & $     -1.07\pm    0.09$\\
J003720.44+401646.3 & $      21.85\pm    0.04$ & $    -0.24\pm    0.04$ & $     -1.07\pm    0.03$ & $    -0.89\pm    0.04$\\
J003720.45+401631.5 & $      21.48\pm    0.03$ & $   -0.03\pm    0.04$ & $    -0.75\pm    0.03$ & $    -0.72\pm    0.04$\\
J003720.56+401621.4 & $      19.88\pm   0.01$ & $      1.73\pm    0.02$ & $     0.85\pm    0.01$ & $    -0.39\pm    0.02$\\
J003720.64+401632.7 & $      21.67\pm    0.03$ & $   -0.07\pm    0.04$ & $     -1.04\pm    0.03$ & $    -0.99\pm    0.04$\\
J003720.89+401634.8 & $      20.67\pm    0.02$ & $    -0.19\pm    0.03$ & $    -0.67\pm    0.02$ & $    -0.53\pm    0.03$\\
J003720.89+401635.2 & $      21.11\pm    0.04$ & $   -0.06\pm    0.07$ & $    -0.88\pm    0.07$ & $    -0.84\pm    0.09$\\
J003721.84+401624.9 & $      22.17\pm    0.07$ & $    -0.36\pm    0.08$ & $    -0.94\pm    0.04$ & $    -0.67\pm    0.07$\\
\hline
\end{tabular}
\end{table*}
\begin{figure}
\begin{center}
\includegraphics[scale=0.26,angle=0]{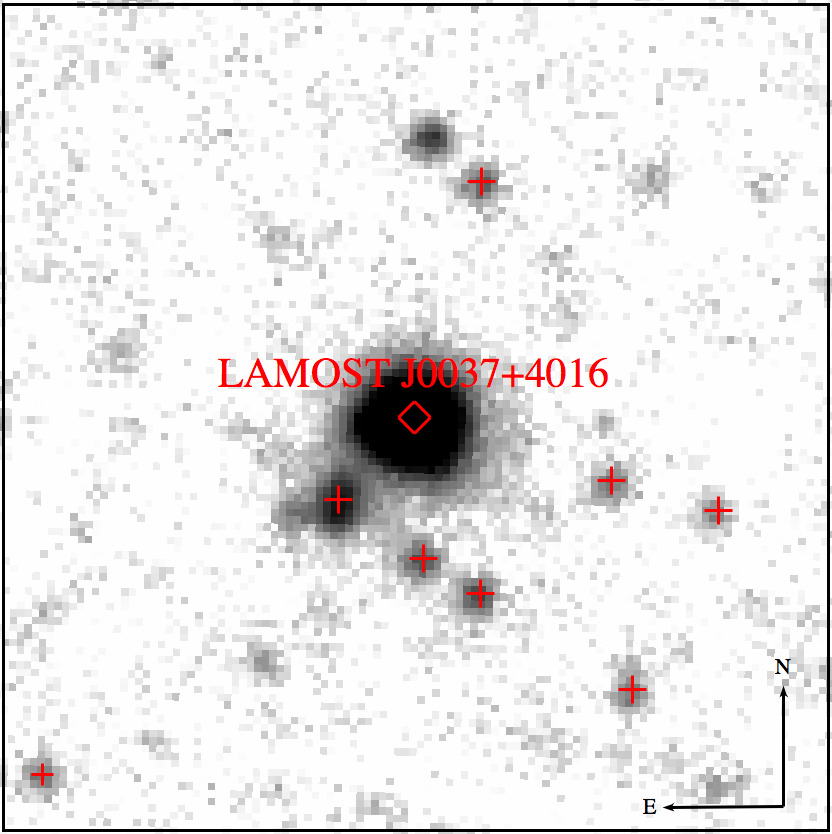}
\caption{ LGGS $U$ band image of LAMOST\,J0037+4016. 
               The whole filed has size of $30$\arcsec\,$\times 30$\arcsec, centred at the location of LAMOST\,J0037+4016 (marked by the red diamond).
               Here the red crosses mark stars with $Q$-index smaller than $-0.6$.}
\end{center}
\end{figure}

\subsection{Near-infrared color-color diagram}
In the upper-left parts of an HR diagram, besides the LBVs, there is another type of massive stars -- B[e] supergiants (B[e]SGs).
In some cases, B[e]SGs are found to show variabilities in magnitude and color/spectral-type similar to LBVs (Zickgraf et al. 1996; Clark et al. 2013).
To further show LAMOST\,J0037+4016 is an LBV rather than a B[e] supergiant, we examine it in a near-infrared color-color diagram.
As pointed out by Oksala e al. (2013), Kraus et al. (2014) and Humphreys et al (2017), LBVs and B[e]SGs have distinct infrared colors, given the larger amount of hot dust surrounding a  B[e]SG.
As Fig.\,2(c) shows, LBVs and B[e]SGs  are indeed well separated in $J - H$ versus $H - K_{\rm s}$ color-color diagram and the LAMOST\,J0037+4016 follows the distribution of Galactic and M31 known LBVs very well.
Therefore, one can conclude that LAMOST\,J0037+4016 is not a B[e]SG but an LBV.

In summary, all existing data, including the light curve, spectra and near-infrared colors, suggest that  LAMOST\,J0037+4016 is {a LBV}.

\section{Physical properties and environments of LAMOST J0037+4016}
\subsection{Physical properties}
In this Section, we derive the physical properties of LAMOST\,J0047+4016 by {its} spectral energy distribution (SED). 
Generally, physical parameters estimated by this way could be ambiguous given the well-known degeneracy between the reddening and temperature.
However, one can break the degeneracy by the fact that LBVs get cooler (brighter) or hotter (dimmer) with almost constant bolometric luminosity (HD94; Sholukhova et al. 2015).

Assuming $T_{\rm eff}^{4}R^{2}$ and the interstellar reddening ($A_{V}$) are both constants, we fit the SED of LAMOST\,J0037+4016 with the multi-band photometric data from LGGS, SDSS and YNAO 2.4\,m, taken {at} different evolutionary phases of the object. 
All the data have fortunately  measurements in an ultraviolet/blue band.
A blackbody is assumed in the SED fitting.
The fits are presented in Fig.\,3.
The best-fit yields luminosity and $A_{V}$ of ($4.42 \pm 1.64$)\,$\times 10^{5} L_{\odot}$ (assuming a distance to M31 of 752\,kpc; Riess et al. 2012) and $0.40 \pm 0.05$, respectively.
The best-fit values of effective temperature and radius are, $10516 \pm 538$\,K and $199 \pm 26$\,$R_{\odot}$ on 2000 October 4 (LGGS), $8393 \pm 571$\,K and $308 \pm 56$\,$R_{\odot}$ on 2002 September 5 (SDSS), and $15235 \pm 1508$\,K and $90 \pm 21$\,$R_{\odot}$ on 2017 January 13 (YNAO 2.4\,m), respectively.
By comparing to the Geneva evolutionary tracks of solar metallicity for an initial rotation velocity of 40\% of the breakup speed (Ekstr{\"o}m et al. 2012), this LBV is found to have an initial mass around 30\,$M_{\odot}$.
A plot of LAMOST\,J0037+4016 and some other well-known LBVs (HD94; de Jager 1998) in $T_{\rm eff}$--$L$ HR diagram {are} also presented in Fig.\,3.
LAMOST\,J0037+4016 falls in the typical region of LBVs in the diagram, within the standard location of quiescent S\,Doradus instability strip (HD94) and the the constant temperature strip of LBVs at their maximum light phase in S\,Doradus variations. 
Finally, as suggested by Humphreys et al. (2016), this new LBV belongs to the less luminous class, that passed through the red supergiant phase and backed again to the left in the HR diagram (HD94).

\subsection{Environments}
As mentioned earlier, LAMOST\,J0037+4016 is the current most distant LBV in M31, located in the south-western corner of M31, with an unexpectedly large projection distance of $\sim 22$\,kpc from the center.
As revealed by {\it Herschel} SPIRE $250$\,$\mu$m (Fritz et al. 2012) and 21\,cm (Braun et al. 2009) images, the site of LAMOST\,J0037+4016 possibly belongs to a faint spiral arm.
To show detailedly the environments of LAMOST\,J0037+4016, the LGGS $U$ band image is shown in Fig.\,4.
In Table\,3, we have also listed all the stars with LGGS photometry {in} $UBV$ bands within $20$\arcsec ($1$\arcsec is about 3.6\,pc at the M31 distance) of LAMOST\,J0037+4016 and most of them (8 out of 10) have $Q$-index smaller than $-0.6$ (earlier than a B2 V or a B8 I; Massey et al. 1998).
The $U$ band image and the $Q$-index of the nearby stars of LAMOST\,J0037+4016 indicate that it may site on a young massive cluster/star forming region and it is the most massive one in this cluster/star forming region.
But further spectroscopic observations of those {close early type} stars {of LAMOST\,J0037+4016} are required and could {further} provide vital clues to the recent debate on the evolutionary status of LBV (e.g. Smith \& Tombleson 2015; Humphreys et al. 2016; Davidson et al. 2016).

\section{Summary}
In this letter, we report the discovery of {a new LBV} --  LAMOST\,J0037+4016 in M31, first confirmed as an M31 supergiant in the LAMOST spectroscopic survey.
This new LBV is located at the south-western corner of M31 with an unexpectedly large projection distance of $\sim 22$\,kpc from the center.

Using archival and newly collected photometric data, we construct $V$ band light curve of LAMOST\,J0037+4016  over a time period of 18 years.
The curve reveals a $V$ band variation amplitude of 1.2\,mag and both outburst and quiescence phases {last} over several years.
From spectra with {the} LAMOST, the YNAO 2.4\,m telescope and the Hale 5.1\,m telescope, we show that LAMOST\,J0037+4016 displays  the characteristics of  a A-type supergiant at epoch close to the outburst phase and a hot early B-type supergiant with weak [Fe II] emission lines at epoch of much dimmer brightness.
From the $J-H$ versus $H-K{\rm s}$ color-color diagram, we find that LAMOST\,J0037+4016 locates in the typical LBV region rather than that of B[e]SGs.
We conclude that all existing data strongly suggest that LAMOST\,J0037+4016 is an LBV.

By SED fitting, we find LAMOST\,J0037+4016 has a luminosity of ($4.42 \pm 1.64$)\,$\times 10^{5} L_{\odot}$ and an initial mass around 30\,$M_{\odot}$, belonging to the less luminosity class of LBV.

 \section*{ \small Acknowledgements} 
 We would like to thank the referee for his/her helpful comments.
 We acknowledge the support of the staff of the Lijiang 2.4 meter and Xinglong 2.16 meter telescopes.
 This work is supported by the National Natural Science Foundation of China U1531244, 11833006, 11811530289, U1731108, 11803029, 11973001 and 11903027 and the Yunnan University grant No.\,C176220100006 and C176220100007.
 
The LAMOST FELLOWSHIP is supported by Special fund for Advanced Users, budgeted and administrated by Center for Astronomical Mega-Science, Chinese Academy of Sciences (CAMS). 
 
The Guoshoujing Telescope (the Large Sky Area Multi-Object Fiber Spectroscopic Telescope, LAMOST) is a National Major Scientific Project built by the Chinese Academy of Sciences. Funding for the project has been provided by the National Development and Reform Commission. LAMOST is operated and managed by the National Astronomical Observatories, Chinese Academy of Sciences.

This research uses data obtained through the Telescope Access Program (TAP), which has been funded by the National Astronomical Observatories, Chinese Academy of Sciences, and the Special Fund for Astronomy from the Ministry of Finance.

Observations obtained with the Hale Telescope at Palomar Observatory were obtained as part of an agreement between the National Astronomical Observatories, Chinese Academy of Sciences, and the California Institute of Technology

\end{document}